# Mathematical Modeling of Business Reopening when Facing SARS-CoV-2 Pandemic: Protection, Cost and Risk

Hongyu Miao[1*], Qianmiao Gao[1], Han Feng[1], Chengxue Zhong[1], Pengwei Zhu[1], Liang Wu[1], Michael D. Swartz[1], Xi Luo[1], Stacia M. DeSantis[1], Dejian Lai[1], Cici Bauer[1], Adriana Pérez[1], Libin Rong[2], David Lairson[3]

[1] Department of Biostatistics and Data Science
School of Public Health, University of Texas Health Science Center at Houston
1200 Pressler St., Houston, TX, 77030

[2] Department of Mathematics
University of Florida
Gainesville, FL 32611

[3] Department of Management, Policy and Community Health
School of Public Health, University of Texas Health Science Center at Houston
1200 Pressler St., Houston, TX, 77030

* Corresponding Author
Hongyu Miao, Ph.D.
Department of Biostatistics and Data Science
School of Public Health, UTHealth
1200 Pressler Street, Houston, TX 77030
Email: hongyu.miao@uth.tmc.edu

## Abstract

The sudden onset of the coronavirus (SARS-CoV-2) pandemic has resulted in tremendous loss of human life and economy in more than 210 countries and territories around the world. While self-protections such as wearing masks, sheltering in place and quarantine policies and strategies are necessary for containing virus transmission, tens of millions of people in the U.S. have lost their jobs due to the shutdown of businesses. Therefore, how to reopen the economy safely while the virus is still circulating in population has become a problem of significant concern and importance to elected leaders and business executives. In this study, mathematical modeling is employed to quantify the profit generation and the infection risk simultaneously from the point of view of a business entity. Specifically, an ordinary differential equation model was developed to characterize disease transmission and infection risk. An algebraic equation is proposed to determine the net profit that a business entity can generate after reopening and take into account the costs associated of several protection/quarantine guidelines. All model parameters were calibrated based on various data and information sources. Sensitivity analyses and case studies were performed to illustrate the use of the model in practice.

# Introduction

Severe Acute respiratory Syndrome Coronavirus 2 (SARS-COV-2) is an ongoing global health threat to every country in the world and has caused significant loss of almost every business entity.[1-3] In the United States alone, as of May 14, 2020, more than 1.3 million confirmed infections and more than 84,000 deaths had been reported since the COVID-19 outbreak, which is associated with a 4.8% drop in gross domestic product (GDP) in the first quarter of 2020. While the development of effective vaccination, treatment and prevention strategies is underway, it is still unknown when such efforts will yield clinical and medical practices effective enough to allow a return to usual economic activity.[4,5] Practicing a variety of stringent quarantine and shutdown policies is an effective way to protect our citizens' health and lives during a pandemic like SARS-CoV-2.[6] However, there are costs that need to be considered. Besides the extraordinary physical and emotional stress and potentially significant medical expenses that COVID patients and their families must cope with, a large number of people have lost (or may lose) their jobs or businesses. This significant financial loss results in a pressure to reopen the economy prior to the availability of effective vaccination and treatment of COVID-19. Elected leaders and business executives as well as employees must address a critical question: how to reopen the economy in the midst of a SARS-CoV-2 pandemic safely?

In this study, we use mathematical modeling techniques to address particular challenges a business entity may face during reopening. In recognition that mathematical modeling results alone cannot stop pandemics,[7] the following behavioral and social guidelines are still strongly suggested for any business entities in planning to reopen:

1. Social distancing (including dinning in restaurants and manufacturing in factories), e.g., the number of workers (and clients or patrons) that a business environment can support while maintaining a 6ft distance between individuals;
2. Mask, glove, and goggle wearing while not alone;
3. Routine sanitization (e.g., entrance, exit, home, workplace, conference room, bathroom, public transportation, door knobs, shared electronic devices);
4. COVID-19 tests accessible to workers who have symptoms or recent exposure to confirmed infections;
5. Deployment of non-contact sensors (e.g., Kinsa smart thermometer) for real-time fever detection and location;
6. Case reporting and quarantine policy;
7. Determination of a maximum time duration of exposure to working environment;
8. Specific equipment (e.g., stronger ventilation system, UV purification system) for aerosol transmission prevention;
9. Employees in non-contact positions remain working from home;

The implementation of each guideline above can protect employees' health and lives but may add an economical cost. Indeed, the main purpose of reopening businesses is to prevent further deterioration of our economy by generating profits and provide incomes that many citizens desperately need. However, for employees in a contact-based position, the risk of COVID infection and transmission may remarkably increase if they return to work. When the infection rate of COVID-19 within a business entity becomes higher than the prevalence in the general population, it may trigger a domino effect and subsequent infections in a broader community originated from this business entity. Therefore, a delicate balance between profit/income generation and infection/transmission prevention must be the ideal. The focus of this modeling strategy is to provide a quantitative tool for investigating whether a business entity can generate positive net profit after reopening while keeping the prevalence of COVID-19 infection within this entity less than or equal to the prevalence in the general population. The prevalence in the general population is assumed to remain under a certain threshold after reopening; otherwise, isolation and quarantine orders may be re-issued by government and reopening would halt.

## Materials and Methods

### Mathematical Model

To our knowledge, there exist only a few studies that have modeled the reopening of a country, a state, a city or a local community,[8-11] and fewer studies have focused on the reopening of a business entity. This modeling work attempts to address the business-reopening problem by considering a transmission model together with a net profit equation.

Borrowing from the classical susceptible-infected-recovered (SIR) modeling framework as in several previous models,[8,12] the following definitions and notations are introduced. Let $N_T$ denote the total number of employees in a business entity, $N_c$ denote the number of employees in a contact-based position, and $N_N$ denote the number of employees in a non-contact position. Employees in a non-contact position are expected to remain working from home (WFH) according to Guideline #9 and thus are excluded from the transmission model. These WFH employees still actively contribute to profit generation and are counted in the net profit equation. In addition, the total number of employees $N_T$ is assumed variable after the reopening due to various reasons, including infection-related death or other factors. Among the $N_c$ employees in a contact-based position, let $S$ denote the number of susceptible employees; $C$ denote the number of asymptomatic, pre-symptomatic, and very-mildly-symptomatic (VMS) carriers (collectively known as silent spreaders); $Q$ denote the number of infected employees on quarantine, under treatment (i.e., confirmed infection) or awaiting test result (i.e., unconfirmed infection); and $D$ denote the number of deaths or resignations due to infection. Note that: i) all infected employees confirmed by virus tests should be quarantined immediately, and any company fellow having a recent contact with the infected person is assumed to also be tested for the virus and initially be self-quarantined (see Guideline #4). The asymptomatic, pre-symptomatic, and very-mildly-symptomatic carriers are the three major categories of the so-called "silent spreaders".[13] While it remains unclear how quickly asymptomatic carriers can transmit the infection, some studies in China (ref) suggested that approximately 25% of those who tested positive without symptoms continued symptomless, and the remaining 75% turned out to be pre-symptomatic.[14] Other ongoing studies suggested that the proportion of asymptomatic infection could be as high as 50%.[15] People with very mild symptoms (e.g., occasional light cough or mild fever) may not recognize the infection but are also fully capable of spreading the disease like the presymptomatic carriers.[16] Finally, the WFH employees may still be infected and then practice at-home quarantine or receive treatment in hospital; for simplicity, instead of introducing another set of equations for characterizing the WFH transmission, the number of infected WFH employees $N_{NQ}$ is assumed to be collected by the business entity on a self-reported basis and directly accounted for in the net profit calculation equation.

After taking silent spreaders into consideration,[17] the proposed model structure is given below:

$$\frac{dS}{dt} = -\beta_O \cdot \kappa \cdot S - \beta_A \cdot \frac{\alpha C}{N} \cdot S - \beta_P \cdot \frac{(1-\alpha)C}{N} \cdot S, \qquad (1)$$

$$\frac{dC}{dt} = \beta_O \cdot \kappa \cdot S + \beta_A \cdot \frac{\alpha C}{N} \cdot S + \beta_P \cdot \frac{(1-\alpha)C}{N} \cdot S - \tau \cdot (1-\alpha)C + \omega \cdot Q, \qquad (2)$$

$$\frac{dQ}{dt} = \tau \cdot (1-\alpha)C - (\gamma + \omega) \cdot Q - \delta \cdot Q, \tag{3}$$

$$\frac{dR}{dt} = \gamma \cdot Q, \tag{4}$$

$$\frac{dD}{dt} = \delta \cdot Q, \tag{5}$$

and the net profit is defined as the following:

$$P_{ntot} = \xi_1 \cdot \xi_7 \cdot \rho \cdot \left((N_N - N_{NQ}) + S + C + R\right) - \sum_{i \neq 1 \text{ or } 7} \xi_i \cdot (S + C + R) - w \cdot Q - w \cdot N_{NQ}, \tag{6}$$

where $N = S + C + R$ denotes the number of employees in contact-based positions who are working on site. The definitions of all model variables and parameters are listed in Table 1, and a diagram (Fig. 1) is also given to illustrate the transmission model mechanisms and assumptions. As suggested in Fig. 1 and Eq. (1), susceptible employees can be infected and become virus carriers by contacting people outside of the business at a rate $\beta_O$, or interacting with asymptomatic carriers within the business at a rate $\beta_A$ and with presymptomatic and VMS carriers within the business at a rate $\beta_P$. Also, $\alpha$ denotes the proportion of asymptomatic carriers among all tested-positive infections, and $\kappa$ denote the probability of transmission via an average number of contacts per day of one person with other persons. In Eq. (2), the infected susceptible become asymptomatic, pre-symptomatic or VMS carriers, and those pre-symptomatic and VMS carriers further progress to symptomatic infections at a rate $\tau$. In Eq. (3), symptomatic infections may recover at a rate $\gamma$ and become immune to virus infection due to, e.g., memory immunity,[18] regress to symptomless carriers (e.g., some patients can test positive and shed viruses after symptoms end[19]) at a rate $\omega$, or die at a rate $\delta$. Also, Eqns. (4) and (5) characterize the dynamics of recovered and death, respectively. In the net profit Eq. (6), $\rho$ denotes the net profit per day generated by healthy WFH, susceptible, and silent carrier employees, $\xi_i$ ($i \neq 1,7$) is the average cost per person per day in a contact-based position associated with the $i$-th protection Guideline ($i = 1, ..., 9$), and $w$ is the average wage per day per person for employees on quarantine or under treatment. It is assumed that employees on quarantine or under treatment still receive their wages given that the typical quarantine or treatment length is 14-28 days even considering them as sick days.

**Data Source and Parameter Values**

The datasets generated from recently published studies and surveys are heterogeneous in terms of population demographics, geospatial characteristics, medical resource availability, treatment regimens, prevention polices and implementation, among others. Extraordinary efforts are needed to integrate such heterogeneous datasets and perform a variety of statistical analyses, which is beyond the scope of this study. Instead, the summary statistics or previously-calibrated parameter values from previous studies and surveys are used as the primary information sources for our model parameter calibration. Note that studies reported different accuracy (in terms of decimals), it is thus difficult to standardize the number of decimals without losing accuracy so we keep the original numbers.

To calibrate the transmission model parameters in Eq. (1)-(5), we started with the case death rate and the recovery time. In the United States, the case fatality rate is currently 5.65% while Omer et al. (2020)[20] previously estimated the infection-related death rate as low as 1.7%. In other countries, the reported case death rates may range from 0.56% (Iceland) to 13.53% (Italy);[20-22] to preserve parameter uncertainty, this wide range is adopted in our calculations. To calculate the mean recovery time, we consider the following observations:[23] i) 8 out of 10 infected persons with symptoms have only mild illness; ii) the average recovery time for mild cases is about 2 weeks; iii) for severe cases, recovery may take up to 6 weeks; iv) the overall recovery rate is between 97% and 99.75%. The conservative mean recovery time for all cases is thus $1/[\left(\frac{0.8}{14} + \frac{0.2}{42}\right) \times 0.97] = 16.65$ days among people with symptomatic infections, which is longer than the reported median hospitalization period of 12 days [24] among survivors. It was also reported that the recovery time for mild cases can be as short as 7 days, therefore, this study assumes that recovery time ranges from 7 to 42 days.[25]

In Eq. (1), for the proportion of asymptomatic carriers (denoted by α), multiple studies have been conducted to estimate this parameter among different populations with different methods.[15,26-29] The range of the estimate of $\hat{\alpha}$ is (13.8%, 75%), and about half of such studies reported a result around 50%. Sun et al. (2020) [26] analyzed 391 cases in Zhejiang Province, China from Jan. 20th, 2020 to Feb. 10th, 2020, and found 54 (13.8%) cases were asymptomatic. Nishiura et al. (2020) [29] estimated α in the evacuated Japanese citizens to be 30.8% (95% CI: 7.7-53.8%). Mizumoto et al. (2020) [15] obtained an estimate 17.9% (95% CI: 15.5-20.2 %) using the data from the Diamond Princess cruise. Kimball et al. (2020) [28] analyzed the results of symptom assessment and SARS-CoV-2 testing in King County, Washington, and found that 56.5% of the tested positive was asymptomatic. Day (2020) [27] suggested that the proportion of asymptomatic infection was between 50% and 75% in northern Italy, and others [30] suggested that $\alpha$ could be between 25% and 50% on Apr. 5th, 2020.[31]

Furthermore, κ is the probability of one employee having contacts with infected carriers outside a business entity, and $\beta_O$ denotes the associated transmission rate. A limited number of studies were found helpful for estimating these two parameters, we thus calibrated such parameter values primarily using the simulation results from Tang et al. (2020).[32] In Tang's study, the daily average contact number was estimated to be 14.78 (SE 0.90) contacts per day per person, and the probability of successful transmission per contact was estimated to be $2.10 \times 10^{-8}$ (SE $1.19 \times 10^{-9}$). According to the definitions of κ and $\beta_O$, we have $\beta_O \cdot \kappa = 14.78 \times 2.10 \times 10^{-8} = 3.10 \times 10^{-7}$ day$^{-1}$, and the range of $\beta_O \cdot \kappa$, $(2.42 \times 10^{-7}, 3.88 \times 10^{-7})$, can be calculated using the standard errors (i.e., $2.42 \times 10^{-7} = (14.78 - 2 \times 0.90) \times (2.10 \times 10^{-8} - 2 \times 1.19 \times 10^{-9})$, and $3.88 \times 10^{-7} = (14.78 + 2 \times 0.90) \times (2.10 \times 10^{-8} + 2 \times 1.19 \times 10^{-9})$).

The transmission rate ($\beta_P$) can be calculated from $\frac{R_0}{recovery\ time}$, where $R_0$ denotes the reproduction number (i.e., the average number of new infections by one carrier) that has been frequently investigated in many SARS-CoV-2 studies.[14,32-41] For instance, according to the study of Liu et al. (2020),[42] the estimated $R_0$ of SARS-CoV-2 in China at the early stage of the pandemic ranges from 1.4 to 6.49 with a mean 3.28 and a median 2.79. Note that the population in Liu et al. (2020) was mostly not aware of the pandemic and used little protections such as mask wearing, hand washing and social distancing. Thus, for the unprotected general population, the estimated transmission rate is 3.28/16.65= 0.197 day$^{-1}$. The corresponding range is between 0.0333 (=1.4/42) and 0.927 (=6.49/7) per day. Note that in the early stage of a pandemic, it can be assumed that most of the cases are pre-symptomatic or mildly-symptomatic such that the parameter value calculated above may be close to the true value of $\beta_P$. Another study by Li et al. (2020)[43] suggested that the transmission rate $\beta_P$ could be 1.12 (95% CI: 1.07-1.18), 0.52 (95% CI: 0.42-0.72), or 0.35 (95% CI: 0.28-0.45) per day by fitting data of different pandemic stages in China. For these reasons, we assumed a range for $\beta_P$ of (0.0333, 1.18) per day.

For the transmission rate of asymptomatic infection ($\beta_A$), very little useful information was found in the scientific literature. Here we borrowed the idea of Li 's work,[43] in which the transmission rate of asymptomatic infection was calculated by multiplying a reduction factor μ with the transmission rate of symptomatic infection. The reduction factor μ was estimated as 0.55 (95% CI: 0.46-0.62), 0.50 (95% CI:

0.37- 0.69) and 0.43 (95% CI: 0.31, 0.61), corresponding to different stages of SARS-CoV-2 outbreak in China. Using the same reduction factor µ in our model, $\beta_A$ can be estimated by $\mu\beta_p$, leading to a mean of $0.197 \times 0.5 = 0.099$ per day and a range from $0.0103 (= 0.333 \times 0.31)$ to $0.814 (= 1.18 \times 0.69)$ per day.

In Eq. (2), $\tau$ is the rate of silent carriers progressing to symptomatic infections. The mean incubation period was previously reported to be 5.2 days with a 95% confidence interval (4.1, 7.0) days,[14,44] or 5.1 days with a 95% confidence interval (2.2, 11.5) days.[45] Based on these results, we estimated $\tau$ as the reciprocal of the incubation period with a mean 1/5.2=0.192 per day and a 95% confidence interval (0.143, 0.244) per day.

In Eq. (3), for the clearance rate of symptomatic infection $\gamma$ and the death rate due to COVID infection $\delta$, multiple studies developed various methods to estimate them.[6,32,46] In Piguillem's method,[6] the calculations were mainly based on the case mortality rate and the recovery time. After substituting 16.65 days for the mean recovery time and 5.65% for the mean death rate aforementioned, we obtain the estimate of $\gamma$ as $(1 - 0.0565)^2/16.65 = 0.0535$ per day, and the death rate $\delta$ as $(1 - 0.0565) \times \frac{0.0565}{16.65} = 0.0032$ per day. The range of $\gamma$ is found to be (0.0178, 0.141) per day, and the range of $\delta$ is $(1.3 \times 10^{-4}, 1.67 \times 10^{-2})$ per day.

In Eq. (2) and (3), we also introduced $\omega$, the rate of regression to carriage, considering the fact that 3.23% of the patients recovered from SARS-CoV-2 infection were tested positive after hospital discharge.[47,48] However, the potential infectivity of these carriers remains unclear. Here $\omega$ is estimated to be 3.23% $\times \hat{\gamma}$=0.0323×0.0535= 0.00173 per day, and its range is between 0.000575 (=0.0178 × 3.23%) and 0.00455 (=0.141×3.23%) per day.

Now consider the additional parameters in the net-profit equation Eq. (6). The average net profit per capita $\rho$ in the U.S. is found to be $400.73 per person per day, which is calculated by dividing Jan 2020 U.S. corporate profit $1908.02 Billion US dollars [49] by 158,714,000 (the number of employed persons in the United States)[50] and then by 30 days. The average weekly wage of one employee in the U.S. is $1,093 per person in the third quarter of Year 2019,[51] so the average daily wage of employee $w$ is $218.60 per day (dividing $1,109 by the number of weekdays 5). In the previous study of Thumstrom and Newbold (2020),[12] GDP loss was considered as one cost of social distancing; i.e., an immediate GDP decline associated with practicing social distancing alone (i.e., house-hold quarantine) in the United States was predicted to be $13.7 - 6.49 = 7.21$ trillion US dollars. The projected GDP of Fiscal Year 2020 is 22.11 trillion US dollars according to the United States Congressional Budget Office (CBO).[52] Therefore, for Guideline #1, the cost of social distancing is the loss of productivity by $7.21 \div 22.11 = 32.6\%$, and thus $\xi_1 = 1 - 32.6\% = 67.4\%$. For Guideline #2, the cost of personal protective equipment (PPE),[53] including surgical mask, gloves, goggle wearing, hand sanitization, and soup, is calculated as $\xi_2$ =$3.60 per person per day [54] under the assumption that each person will consume two surgical masks per day, two pairs of gloves per day, and one goggles. According to the Personal Protective Equipment (PPE) guideline from Perdue University,[55] goggles can last for years if kept clean by using mild soap and water, and if stored in a protected, dry, and temperate storage location. So the use life of goggles is much longer than our typical setting for simulation time window length (i.e., 100 days) and the cost of one goggles per day is $5 \div 100 = $0.05 per day for simplicity. The detailed costs of each PPE item as well as hand sanitization and soap can be found from online resources.[56,57] For Guideline #3, the cost of routine sanitization (e.g., cleaning the workplace, bathroom, and shared electronic devices) is calculated as $\xi_3 = $10.45 per environmental service staff per day.[58] Specifically, in the work of White (2019), 11 hospitals with a total of 1,700 environmental service staffs, they found to spend $11,308 per week to maintain the hygiene by consuming disinfection products. We thus calculate $\xi_3$ using the following equation:

$$\xi_3 = \frac{cost\ of\ disinfection\ products}{\left(\frac{Total\ number\ of\ enviromental\ services\ staffs}{Number\ of\ hospitals}\right)} \div 7\ days.$$

For Guideline #4, the current policy dictates that the test of COVID is free to the business entity ($\xi_4 =$ $0 per person per day). For Guideline #5, as the current market prices of a non-contact sensor may range from $50 to $100, and one non-contact sensor is required for each worksite. With the assumption of having 100 employees per worksite, the cost of deploying non-contact sensor $\xi_5$ is $0.005 to $0.01 per person per day for a 100-day time horizon. We assume that Guideline #6 does not cost any money ($\xi_6 =$ $0 per person per day). For Guideline #7, the cost is proportional to the current maximum working hours (i.e., $\xi_7 = \frac{current\ maximum\ working\ hours}{regular\ working\ hours}$). In our model, the current maximum working hours is assumed to be 70% of the regular working hours. For Guideline #8, referring to the work of Chen (2013),[54] the cost of deploying specific equipment for reducing aerosol transmission such as ultraviolet germicidal irradiation (UVGI) and high-efficiency particular air (HEPA) filtration are $182.37 and $136.78 per person per year, respectively, $\xi_8$ is thus equal to $0.874 per person per day via dividing the total cost of the aforementioned equipment by 365 days. Finally, we assume that implementing Guideline #9 does not incur any cost ($\xi_9 =$ $0 per person per day).

All the parameter definitions, values and ranges are summarized in Table 1. However, it should be stressed that the parameter values calibrated above are for heterogeneous populations. More importantly, at the early stage of the pandemic, the estimates of certain transmission model parameters (e.g., the reproduction number and transmission rates) are expected to be larger due to the absence of protection policies and self-protection awareness; and the transmission rates are expected to have a notable drop after the implementation of various protection and quarantine strategies (PQS). Such a hypothesis is supported by several recent studies, which showed that the overall transmission rates may decrease by 2.1 to 3.2 fold after implementing PQS. Also, the study of Seto et al. (2003) [59] quantified the odds ratios of SARS infection as 13 (95% CI: 3, 60), 2 (95% CI: 0.6, 7), or 5 (95% CI: 1, 19), corresponding to the use of masks, gloves or hand-washing, respectively. Thus, the values of certain model parameters (e.g., transmission rates) will be different from (e.g., smaller than) the values calibrated in this section after implementing Guidelines #1-9, which will be elaborated in the result Section.

**Implementation and Computing Configuration**

All the computing codes were implemented in MATLAB® (MathWorks, Natick, MA), and the ordinary differential equation (ODE) solver ode15s was employed for solving the transmission model numerically. The relative error tolerance was set at $10^{-7}$, the absolute error tolerance at $10^{-7}$, and the maximum step size at $10^{-2}$.

## Results

### Sensitivity Analysis

We evaluated the local sensitivity of the transmission model in Eq. (1)-(5). Specifically, we evaluated the changes in the model outcome variables (i.e. S, C, Q, R, D) corresponding to a 1% change in one parameter value, with the other seven parameter values fixed at their default values as in Table 1. Initial values for the sensitivity analysis were set as 299 susceptible and 1 silent carrier which approximated the proportion of infections in the U.S. general population as reported by May 5, 2020. The initial numbers of quarantined, recovered, and death were all set as 0 for simplicity. The results were visualized in Figure 2a, showing that the transmission model was most sensitive to the transmission rate of pre-symptomatic and VMS infections ($\beta_P$) and the rate of progression to symptomatic infection ($\tau$), and least sensitive to the product $\beta_O \cdot \kappa$. Even for the most sensitive parameters, a 1% change in parameters $\beta_P$ or $\tau$ resulted in a change of less than 5 persons per day in the outcomes throughout a prediction period of 200 days. At the end of the 200 days period, all changes in the outcomes reached a plateau or tended to diminish. The percentage changes in the model outcomes corresponding to 1% change in each parameter value are shown in Figure 2b. One percent change in $\beta_P$ or $\tau$ corresponded to less than 1.5% change in the outcomes. These observations together suggested that the transmission model was not locally sensitive to parameter value changes and could make robust predictions.

To evaluate the global sensitivity of this model, the Partial Spearman Rank Correlation Coefficient (PRCC) method [60] was employed. The model parameters were randomly sampled 100 times using the Latin hypercube sampling technique over uniform distribution. The range of the uniform distribution for the global sensitivity for each model parameter is reported in Table 1. Figure 3, shows the plot of the PRCC values (next to the y axis in the figure) between model parameters and ODE outputs against time. As suggested in many previous studies, an absolute PRCC value greater than 0.4 was deemed as practically significant.[61] For the susceptible population ($S$), the corresponding subfigure showed that the transmission rates $\beta_A$ and $\beta_P$ were strongly and negatively ($< -0.7$) correlated with $S$. This result was expected as the higher the transmission rates, the smaller the number of the susceptible persons. Please note that the percentage of asymptomatic carriers ($\alpha$) was positively correlated with $S$ initially ($> 0.4$) and then became negatively correlated with $S$ at the end ($< -0.6$). The initial positive correlation between $S$ and $\alpha$ should not be interpreted as that $S$ will increase when $\alpha$ increases, but that $S$ will decrease less when $\alpha$ increases. This will happen when $\beta_A \cdot \alpha$ is less than $\beta_P \cdot (1 - \alpha)$. Around the end of the time window of the simulation, a larger $\alpha$ corresponded to a smaller $N = S + C + R$ due to the increase in death, and thus $S$ will decrease more given a smaller $N$ in the denominator at the righthand side of Eq. (1). From all the subfigures, the transmission rates $\beta_A$ and $\beta_P$ were always found strongly correlated with the outcomes while it was not the case for $\beta_O \cdot \kappa$, the value of which was too small to have a substantial impact on the outcomes. In addition, parameters $\alpha$, $\gamma$ and $\tau$ were also found strongly correlated with the outcomes, and interventions like more effective drug treatment or vaccination would affect such parameter values.

**Case Study**

Business executives are strongly recommended to follow the guidelines listed in the Background Section to prevent the reopening from causing any exacerbation of the ongoing pandemic. However, practical difficulties may arise due to, e.g., insufficiency of budget or medical recourses such that only some of the guidelines will be implemented by employers. Through this case study, we illustrated the use of the transmission model and the cost equation in different scenarios to evaluate the feasibility of reopening. Note that our simulation results were obtained under many assumptions and subject to both model structure and parameter value uncertainty; therefore, decisions of business executives should not be made solely based on the results presented in this section.

In this case study, we focused on four scenarios: I) none of the nine Guidelines was constantly implemented (baseline scenario); II) all Guidelines 1-9 were constantly implemented; III) Guidelines 1, 2, 3, 4, 6, 9 were constantly implemented; IV) Guidelines 1, 3, 4, 6, 9 were constantly implemented. Scenarios I was the baseline scenario, corresponding to complete devalue of infection risk. Across scenarios II-IV, Guidelines 1, 3, 4, 6, 9 were assumed to be always implemented given the necessity and indispensability of these five Guidelines to business reopening. Scenario III was designed to be less restrictive than Scenario II, considering that some business entities might not have the budget to deploy non-contact sensors for real-time fever detection, reduce the number of working hours, or acquire specific equipment for aerosol transmission prevention. In Scenario IV, the use of PPE was further dropped to account for the possible shortage of such materials on the market. The implementation of behavioral and social practice guidelines in Scenarios II-IV led to changes in the values of these three, $\beta_O \cdot \kappa$, $\beta_A$, and $\beta_P$, parameters in our transmission model. The other model parameters were assumed not directly affected by Guidelines 1-9. For instance, the recovery rate $\gamma$ primarily depends on subject-specific immunity, the availability, affordability of effective medical intervention and health care, instead of behavioral and social practice patterns. For these reasons, we adjusted the values of $\beta_O \cdot \kappa$, $\beta_A$, and $\beta_P$ for scenarios II-IV, respectively, as shown in Table 2. For simplicity, we assumed that the effects of different guidelines were independent and remained constant throughout the entire simulation time window. We also assumed that after reopening, every employee rigorously followed the guidelines and immediately reported their symptoms or infections once identified.

According to the study of Koo et al. (2020), [62] the practice of social distancing together with disease testing, reporting and quarantine policy (Guidelines 1, 4, and 6) could reduce the number of infections by 78.20% (IQR: 59.0-94.4%) compared with the baseline scenario when R$_o$ was 2.5. That is, the transmission rates might drop to 1-78.2%=21.80% of their baseline values after implementing Guidelines 1, 4 and 6. To quantify the effects of wearing mask, glove, goggle and hand washing (Guideline 2) on infection transmission, we adopted the results from Seto et al. (2003) [59] and Yin et al. (2004) [63] for severe acute respiratory syndrome (SARS). In their results, the odds ratio for mask wearing was 13 (95% CI: 3-60, attack rate $\frac{11}{83} = 13.25\%$), for glove wearing was 2 (95% CI: 0.6-7, attack rate $\frac{9}{133} = 6.77\%$), for hand-washing was 5 (95% CI: 1-19, attack rate $\frac{3}{17} = 17.65\%$), and for goggles wearing was $\frac{1}{0.2} = 5$ (95% CI: $2.44 - 10$, attack rate 61.50%). Correspondingly, the transmission rates might drop to 8.83% (95% CI: 1.92-36.56%) for mask wearing, 48.85% (95% CI: 15.17-100%) for glove wearing, 39.37% (95% CI: 22.40-64.33%) for goggle wearing, and 23.91% (95% CI: 6.32-100%) for hand washing. Thus, due to the implementation of Guideline 2, the transmission rates might drop to $8.83\% \times 48.85\% \times 39.37\% \times 23.91\% = 0.41\%$ of their baseline values. Zhang et al. (2018) [64] investigated influenza A transmission in a student office setting, which showed that the infection risk could reduce by 2.14% with implementation of routine sanitization. Routine sanitation is expected to reducethe transmission rates to $1 - 2.14\% = 97.86\%$ of their baseline values after implementing Guideline 3. The implementation of Guideline 5 would improve the implementation of Guidelines 4 and 6 so its effect on infection transmission was not explicitly quantified in this study. For Guideline 7, we assumed a linear relationship between the number of working hours and the infection risk. The average working hours per day after reopening was assumed to be 70% of the regular working hours. Correspondingly, the transmission rates might drop to 70% (range 50-100%) with Guideline 7. For Guideline 8, according to the report of Mendell et al (2002),[65] a 10% to 14% reduction in communicable respiratory infections might result from improved work environments. Taking 12% as the median, we assumed that the transmission rates might drop to $100\% - 12\% = 88\%$ (range 86-90%) with implementing Guideline 8. Furthermore, the transmission rates among employees in contact-based positions were not affected by Guideline 9. Finally, in scenarios II, III, and IV, the transmission rates dropped to $21.80\% \times 0.41\% \times 97.86\% \times 70\% \times 88\% = 0.05\%$, $21.80\% \times 0.41\% \times 97.86\% = 0.09\%$ and $21.80\% \times 97.86\% = 21.33\%$ of their baseline values, respectively. See Table 2 for the adjusted transmission rates.

All the simulation results in this section were generated using the same set of initial values for simplicity. Taking the DELL$^{TM}$ center at Austin, Texas as an example, the total number of active workers after reopening could be around 14,000. The ratio between the on-site workers and the WFH workers was assumed as 2:1 (i.e., 9,333 on-site workers and 4,667 WFH workers). On the first day of reopening (day 0), the proportion of silent carriers among all the 14,000 employees was approximated by the proportion of infections in the general population of U.S. estimated using the number of reported cases as of May 5, 2020. The proportion of workers that had recovered from the infection and thus acquired immunity (referred as "recovered" in our model) were also estimated from the reported number of recoveries of the general U.S. population. The numbers of quarantined employees and death on day 0 were assumed to be 0. We conducted the simulations for 200 days for Scenario I and 100 days for Scenarios II-IV to verify the short-term feasibility of reopening.

As shown in Figures 4-5 for Scenario I, without implementing any of the guidelines, reopening merely led to a large number of infections and deaths (230 deaths by the last day, 649 quarantined and 422 silent carriers at their peaks in Figure 4). The prevalence of infections within the business entity exceeded that in the general population throughout the 200 days time window, and peaked at 114 infections per 1000 people, which was 30 folds higher than that of the general population (Figures 5a). Even by the end of the 200 days, the within-entity prevalence was 8 folds higher (32.9 cases per 1,000 persons) than the population average. Also, even after we extended the simulation time window to 200 days, the transmission model still did not reach its steady states, suggesting a less predictable risk of disease transmission. In short, despite that the net profit remained positive (Figure 5b), since the prevalence in workplace was constantly (much) higher than that in the general population, reopening turned out to be infeasible in Scenario I.

If a business entity strictly followed all guidelines 1-9 as in Scenario II, the number of infections and deaths reduced remarkably, as shown in Figures 6-7 (2 deaths among 9,333 onsite workers, at most 16 quarantined and 35 silent carriers as in Figure 6). Figure 6 also showed that the outcomes of the transmission model reached a plateau towards the end of the 100-days time window. In other words, the infection could be contained in this scenario, with the number of carriers and the number of the quarantined reaching and staying at 0 by day 48 and day 92, respectively. According to Figure 7a, under the assumption that the prevalence of infections within the business entity was the same as in the general population at the beginning of reopening, the within-business prevalence dropped under the general population prevalence immediately (on day 2), and continuously decreased down to 0.035 infections per 1,000 people by the last day of simulation. This within-business prevalence was 104 fold lower than the general population prevalence reported in other studies. The business also attained higher and more stable net profit than in Scenario I (Figure 7b).

In Scenario III, guidelines 5, 7, 8 were skipped and the transmission rate values changed accordingly (see Table 2); however, the simulation results were surprisingly similar to those in Scenario II during the 100-day time window (Figures 8-9). Specifically, the numbers of deaths, the quarantined and silent carriers were nearly the same as those in Scenario II; also, it took approximately the same amount of time for the numbers of carriers and the quarantined to drop to 0 (Figure 8). The nearly same predicted population trajectories resulted in a nearly same prediction on net profit (Figure 9). To confirm such results, additional local sensitivity analysis was conducted at the parameter values in Scenarios II and III, suggesting that the simulation outputs were not sensitive to parameter value changes (e.g., a change of less than 0.1 in all the five output variables corresponding to 1% parameter value change).

After a business entity further dropped guideline 2 (Mask, glove, and goggle wearing while not alone) in Scenario IV, our model predicted that the spread of infections did not become uncontrollable within 100 days but associated with a higher cost (Figures 10-11). While the initial within-business prevalence was the same as the general population prevalence, the workplace quickly became a "hot spot" of infection spreading, and a prevalence much higher than the population average was reached (Figure 11a). The within-business prevalence continued to stay above the general population average for 11 days, and peaked at a level of 4 infections per 1,000 people on day 5. The results suggested that reopening should stop to prevent this entity from developing into a source of infection and posing significant risk on its workers as well as their close social contacts. Note that the within-business prevalence dropped to 0.09 infections per 1,000 people by the end of the 100-day time window (Figure 11a), and the numbers of deaths and the quarantined were controlled under 3 and 20 among 9,333 onsite workers, respectively. Business executives should not rely on such optimistic predictions and underestimate the infection risk for two reasons. First, constant parameter values were used in our simulations, which were not capable of capturing every possible time-varying characteristic of disease transmission over time (i.e., parameter values could be time-varying instead of being constant). Second, our simulation was performed under the assumption that all workers strictly followed the selected guidelines. In reality, it is unlikely that every single employee would faithfully stick to such rules and guidelines daily and constantly over time. Moreover, Figure 11a suggested that the use of PPE was very important at the early stage of reopening. In Scenarios II and III, with the use of PPE, we did not observe the rapid increase of prevalence since the beginning of reopening; however, in Scenario IV without the protection of PPE, the within-business prevalence started to increase and exceed the population average on day one. This result was consistent with the recent study by Kai et al (2020), [66] which demonstrated the significant effect of universal use of facial masks (e.g., at least 80% population wear masks) on impeding the spread of infections.

## Discussion

SARS-CoV-2 struck the whole world since 2019 and caused significant loss of human lives and economy. While various lockdown, quarantine and isolation rules and polices worked effectively to control COVID infection transmission, numerous businesses were closed and tens of millions people lost their jobs. As of May 20, 2020, all the states in the U.S. strategically moved towards gradual reopening to save the economy. However, given that neither effective drug treatments nor vaccines were available, the risk of infection spreading within a business entity became a key issue that any business executive had to consider. In this study, an ODE-based transmission model was developed together with a net profit equation to quantitatively evaluate the trade-off among profit, cost, and infection risk within a business entity after reopening. Model parameter values were calibrated from heterogeneous sources to enable computer simulations. Both local and global sensitivity analyses were conducted to understand our model behavior and result robustness. Finally, a case study assessed scenarios, in which different combinations of behavioral and social practice guidelines were implemented. The simulation results suggested that infection transmission was controllable within a business entity and a positive net profit could be generated after reopening only if a combination of selected guidelines were implemented. Also, our results suggested that the use of PPE could be significantly important at the early stage of reopening to prevent infection spreading.

We also recognize a number of limitations of this study. First, the mathematical model and the net profit equation were developed under multiple assumptions. While many of those assumptions are commonly adopted by related professional communities (e.g., epidemiology or mathematical modeling), some assumptions were introduced due to the lack of accurate and/or complete information (e.g., asymptomatic infection transmission rate). With additional efforts invested in future SARS-CoV-2 related research , we expect that informative and high-quality data will become available such that less assumptions are needed and more accurate results can be generated by our approach. Second, while we have compiled information from many different sources to calibrate the possible ranges of model parameter values, the parameter uncertainty may not be completely characterized by such parameter ranges due to, e.g., the heterogeneity in population demographics, health conditions, behavioral patterns, and social networks. Third, although the costs associated with guidelines #1-9 were included in the net profit calculation, our cost estimation is subject to market fluctuations and many other (unpredictable) factors. It is thus suggested for users of our model to fine tune parameter values upon available business-specific data or information. Finally, in this study, herd immunity and vaccination strategies were not considered because they were not available as of the current moment. Considering the active research on SARS-CoV-2 vaccine and drug development, the availability of effective vaccines and medications can be expected and it should be taken into consideration in our modeling work at some point.

In summary, this modeling work provides a quantitative tool for decision-makers to explore and evaluate the business reopening option in the midst of COVID pandemics, and it can be extended to similar scenarios (e.g., outbreak of unknown or new virus) by re-calibrating related parameter values. We expect further research efforts in this direction to better prepare for possible strikes of infectious diseases in the future.

## Acknowledgement

This work was partially supported by NSF grant DMS-1620957 (HM), Weatherhead Foundation (DL), and NSF grant DMS-1758290 (LR). Also, thank Dr. Eric Boerwinkle for motivating this study and sharing constructive comments, and thank Drs. Wenyaw Chan and Hulin Wu for insightful discussions and feedbacks.

## References

1. Atkeson A. What Will Be the Economic Impact of COVID-19 in the US? Rough Estimates of Disease Scenarios. *National Bureau of Economic Research Working Paper Series.* 2020;No. 26867.
2. Anderson RM, Heesterbeek H, Klinkenberg D, Hollingsworth TD. How will country-based mitigation measures influence the course of the COVID-19 epidemic? *The Lancet.* 2020;395(10228):931-934.
3. McKibbin WJ, Fernando R. The Global Macroeconomic Impacts of COVID-19: Seven Scenarios. *CAMA Working Paper No 19/2020.* March 2, 2020;Available at SSRN: https://ssrn.com/abstract=3547729 or http://dx.doi.org/10.2139/ssrn.3547729.
4. Wang Y, Zhang D, Du G, et al. Remdesivir in adults with severe COVID-19: a randomised, double-blind, placebo-controlled, multicentre trial. *The Lancet.* 2020.
5. Williamson BN, Feldmann F, Schwarz B, et al. Clinical benefit of remdesivir in rhesus macaques infected with SARS-CoV-2. *bioRxiv.* 2020:2020.2004.2015.043166.

6. Piguillem F, SHI L. Optimal Covid-19 Quarantine and Testing Policies. *CEPR Discussion Paper No DP14613.* 2020;Available at SSRN: https://ssrn.com/abstract=3594243
7. Huremović D. Social Distancing, Quarantine, and Isolation. In: Huremović D, ed. *Psychiatry of Pandemics: A Mental Health Response to Infection Outbreak.* Cham: Springer International Publishing; 2019:85-94.
8. Song B, Hei X. Models and Strategies on Reopening Lockdown Societies Due to COVID-19. *OSF Preprints.* 2020;doi:10.31219/osf.io/umtvh.
9. Bhatia R, Klausner J. A Step-by-Step Plan to Reopen California. *https://thebolditaliccom/covid-19-next-steps-for-california-c01632c8e8b6.* 2020.
10. Aboelkassem Y. COVID-19 pandemic: A Hill type mathematical model predicts the US death number and the reopening date. *medRxiv.* 2020:2020.2004.2012.20062893.
11. Wang X, Tang S, Chen Y, Feng X, Xiao Y, Zongben X. When will be the resumption of work in Wuhan and its surrounding areas during COVID-19 epidemic? A data-driven network modeling. *SCIENTIA SINICA Mathematica.* 2020;https://doi.org/10.1360/SSM-2020-0037.
12. Thunstrom L, Newbold S, Finnoff D, Ashworth M, Shogren JF. The Benefits and Costs of Using Social Distancing to Flatten the Curve for COVID-19. *Forthcoming Journal of Benefit-Cost Analysis.* April 14, 2020;http://dx.doi.org/10.2139/ssrn.3561934.
13. Victor J-M. COVID-19 : HOW TO FIND SILENT SPREADERS ? In:2020.
14. Li Q, Guan X, Wu P, et al. Early Transmission Dynamics in Wuhan, China, of Novel Coronavirus–Infected Pneumonia. *New England Journal of Medicine.* 2020;382(13):1199-1207.
15. Mizumoto K, Kagaya K, Zarebski A, Chowell G. Estimating the asymptomatic proportion of coronavirus disease 2019 (COVID-19) cases on board the Diamond Princess cruise ship, Yokohama, Japan, 2020. *Eurosurveillance.* 2020;25(10):2000180.
16. Bai Y, Yao L, Wei T, et al. Presumed Asymptomatic Carrier Transmission of COVID-19. *JAMA.* 2020;323(14):1406-1407.
17. Chisholm RH, Campbell PT, Wu Y, Tong SYC, McVernon J, Geard N. Implications of asymptomatic carriers for infectious disease transmission and control. *Royal Society Open Science.* 2018;5(2):172341.
18. Dijkstra J, Hashimoto K. Expected immune recognition of COVID-19 virus by memory from earlier infections with common coronaviruses in a large part of the world population [version 1; peer review: awaiting peer review]. *F1000Research.* 2020;9(285).
19. Wölfel R, Corman VM, Guggemos W, et al. Virological assessment of hospitalized patients with COVID-2019. *Nature.* 2020.
20. Omer SB, Malani P, del Rio C. The COVID-19 Pandemic in the US: A Clinical Update. *JAMA.* 2020;323(18):1767-1768.
21. https://ourworldindata.org/coronavirus#what-do-we-know-about-the-risk-of-dying-from-covid-19.
22. https://www.cdc.gov/coronavirus/2019-ncov/cases-updates/cases-in-us.html.
23. https://www.webmd.com/lung/covid-recovery-overview#1.
24. Guan W-j, Ni Z-y, Hu Y, et al. Clinical Characteristics of Coronavirus Disease 2019 in China. *New England Journal of Medicine.* 2020;382(18):1708-1720.
25. https://www.houstonmethodist.org/blog/articles/2020/apr/recovering-from-coronavirus-what-to-expect-during-and-after-your-recovery/.
26. Sun WW, Ling F, Pan JR, et al. Epidemiological characteristics of 2019 novel coronavirus family clustering in Zhejiang Province. *Zhonghua Yu Fang Yi Xue Za Zhi [Chinese Journal of Preventive Medicine].* 2020;54:E027-E027.
27. Day M. Covid-19: identifying and isolating asymptomatic people helped eliminate virus in Italian village. *BMJ.* 2020;368:m1165.
28. Kimball A, Hatfield KM, Arons M, others a. Asymptomatic and Presymptomatic SARS-CoV-2 Infections in Residents of a Long-Term Care Skilled Nursing Facility — King County, Washington. *MMWR Morb Mortal Wkly Rep March 2020.* 2020;69:377–381.

29. Nishiura H, Kobayashi T, Miyama T, et al. Estimation of the asymptomatic ratio of novel coronavirus infections (COVID-19). *International Journal of Infectious Diseases.* 2020;94:154-155.
30. https://www.augustahealth.com/health-focused/covid-19-asymptomatic-carriers-and-antibody-tests.
31. Gudbjartsson DF, Helgason A, Jonsson H, et al. Spread of SARS-CoV-2 in the Icelandic Population. *New England Journal of Medicine.* 2020.
32. Tang B, Wang X, Li Q, et al. Estimation of the Transmission Risk of the 2019-nCoV and Its Implication for Public Health Interventions. *Journal of Clinical Medicine.* 2020;9(2):462.
33. Wu JT, Leung K, Leung GM. Nowcasting and forecasting the potential domestic and international spread of the 2019-nCoV outbreak originating in Wuhan, China: a modelling study. *The Lancet.* 2020;395(10225):689-697.
34. Shen M, Peng Z, Xiao Y, Zhang L. Modelling the epidemic trend of the 2019 novel coronavirus outbreak in China. *bioRxiv.* 2020:2020.2001.2023.916726.
35. Liu T, Hu J, Kang M, et al. Transmission dynamics of 2019 novel coronavirus (2019-nCoV). *bioRxiv.* 2020:2020.2001.2025.919787.
36. Read JM, Bridgen JR, Cummings DA, Ho A, Jewell CP. Novel coronavirus 2019-nCoV: early estimation of epidemiological parameters and epidemic predictions. *medRxiv.* 2020:2020.2001.2023.20018549.
37. Majumder M, Mandl KD. Early Transmissibility Assessment of a Novel Coronavirus in Wuhan, China. *Available at SSRN: https://ssrncom/abstract=3524675 or http://dxdoiorg/102139/ssrn3524675.* 2020.
38. Cao Z, Zhang Q, Lu X, et al. Estimating the effective reproduction number of the 2019-nCoV in China. *medRxiv.* 2020:2020.2001.2027.20018952.
39. Zhao S, Lin Q, Ran J, et al. Preliminary estimation of the basic reproduction number of novel coronavirus (2019-nCoV) in China, from 2019 to 2020: A data-driven analysis in the early phase of the outbreak. *International Journal of Infectious Diseases.* 2020;92:214-217.
40. Imai N, Cori A, Dorigatti I, et al. Report 3: Transmissibility of 2019-nCov. *COVID 19 Resources, Imperial College London COVID-19 reports.* 2020.
41. Riou J, Althaus CL. Pattern of early human-to-human transmission of Wuhan 2019 novel coronavirus (2019-nCoV), December 2019 to January 2020. *Eurosurveillance.* 2020;25(4):2000058.
42. Liu Y, Gayle AA, Wilder-Smith A, Rocklöv J. The reproductive number of COVID-19 is higher compared to SARS coronavirus. *Journal of Travel Medicine.* 2020;27(2).
43. Li R, Pei S, Chen B, et al. Substantial undocumented infection facilitates the rapid dissemination of novel coronavirus (SARS-CoV-2). *Science.* 2020;368(6490):489-493.
44. Linton NM, Kobayashi T, Yang Y, et al. Incubation Period and Other Epidemiological Characteristics of 2019 Novel Coronavirus Infections with Right Truncation: A Statistical Analysis of Publicly Available Case Data. *Journal of Clinical Medicine.* 2020;9(2):538.
45. Lauer SA, Grantz KH, Bi Q, et al. The Incubation Period of Coronavirus Disease 2019 (COVID-19) From Publicly Reported Confirmed Cases: Estimation and Application. *Annals of Internal Medicine.* 2020;172(9):577-582.
46. Zhou F, Yu T, Du R, et al. Clinical course and risk factors for mortality of adult inpatients with COVID-19 in Wuhan, China: a retrospective cohort study. *The Lancet.* 2020;395(10229):1054-1062.
47. Lan L, Xu D, Ye G, et al. Positive RT-PCR Test Results in Patients Recovered From COVID-19. *JAMA.* 2020;323(15):1502-1503.
48. Xing Y, Mo P, Xiao Y, Zhao O, Zhang Y, Wang F. Post-discharge surveillance and positive virus detection in two medical staff recovered from coronavirus disease 2019 (COVID-19), China, January to February 2020. *Eurosurveillance.* 2020;25(10):2000191.
49. https://tradingeconomics.com/united-states/corporate-profits.

50. http://www.dlt.ri.gov/lmi/laus/us/usadj.htm.
51. https://www.bls.gov/regions/southwest/news-release/countyemploymentandwages_texas.htm.
52. https://www.cbo.gov/publication/56073#_idTextAnchor148.
53. Wang Q, Shi N, Huang J, et al. Effectiveness and cost-effectiveness of public health measures to control COVID-19: a modelling study. *medRxiv.* 2020:2020.2003.2020.20039644.
54. Chen SC, Liao CM. Cost-effectiveness of influenza control measures: a dynamic transmission model-based analysis. *Epidemiology and Infection.* 2013;141(12):2581-2594.
55. https://www.chem.purdue.edu/chemsafety/Training/PPETrain/ppetonline.htm.
56. https://www.shopp.org.
57. https://www.amazon.com.
58. White NM, Barnett AG, Hall L, et al. Cost-effectiveness of an Environmental Cleaning Bundle for Reducing Healthcare-associated Infections. *Clinical Infectious Diseases.* 2019.
59. Seto WH, Tsang D, Yung RWH, et al. Effectiveness of precautions against droplets and contact in prevention of nosocomial transmission of severe acute respiratory syndrome (SARS). *The Lancet.* 2003;361(9368):1519-1520.
60. Marino S, Hogue IB, Ray CJ, Kirschner DE. A methodology for performing global uncertainty and sensitivity analysis in systems biology. *Journal of Theoretical Biology.* 2008;254(1):178-196.
61. Overholser BR, Sowinski KM. Biostatistics Primer: Part 2. *Nutrition in Clinical Practice.* 2008;23(1):76-84.
62. Koo JR, Cook AR, Park M, et al. Interventions to mitigate early spread of SARS-CoV-2 in Singapore: a modelling study. *The Lancet Infectious Diseases.*
63. Yin WW, Gao LD, Lin WS, et al. Effectiveness of personal protective measures in prevention of nosocomial transmission of severe acute respiratory syndrome. *Zhonghua Liu XIng Bing Xue Za Zhi.* 2004;25(1):18-22.
64. Zhang N, Li Y. Transmission of Influenza A in a Student Office Based on Realistic Person-to-Person Contact and Surface Touch Behaviour. *International Journal of Environmental Research and Public Health.* 2018;15(8):1699.
65. Mendell MJ, Fisk WJ, Kreiss K, et al. Improving the Health of Workers in Indoor Environments: Priority Research Needs for a National Occupational Research Agenda. *American Journal of Public Health.* 2002;92(9):1430-1440.
66. Kai D, Goldstein GP, Morgunov A, Nangalia V, Rotkirch A. Universal masking is urgent in the covid-19 pandemic: Seir and agent based models, empirical validation, policy recommendations. *arXiv preprint arXiv:200413553.* 2020.

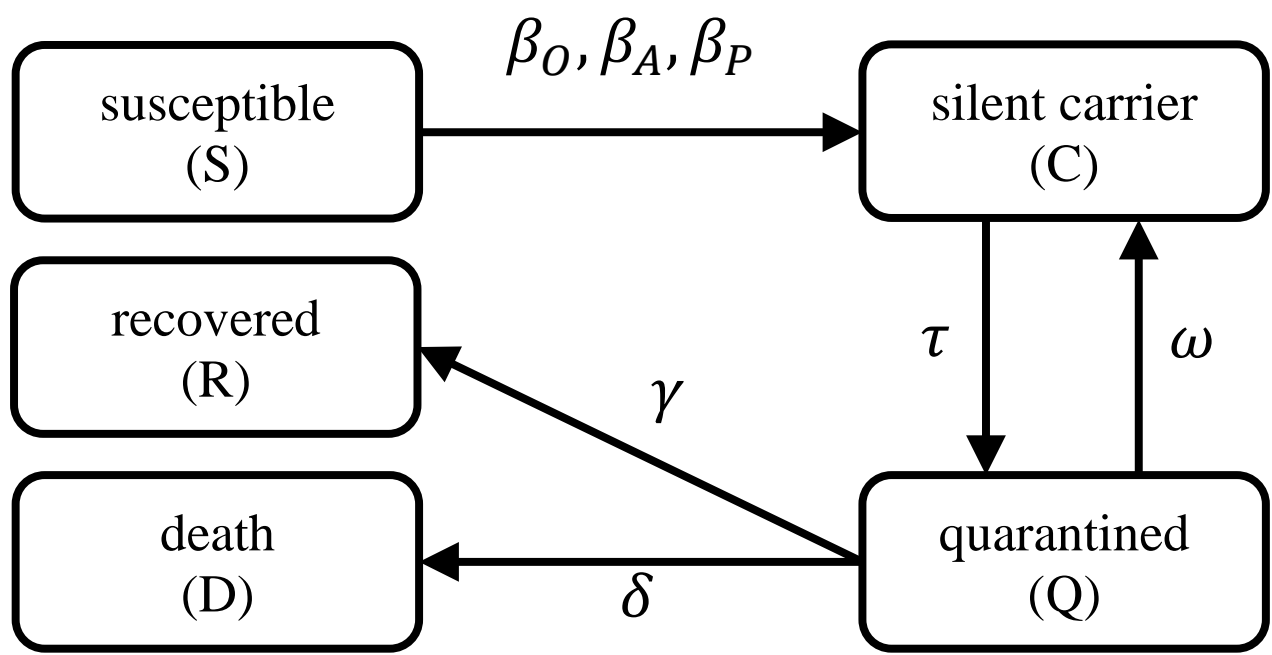

Figure 1. Transmission model diagram.

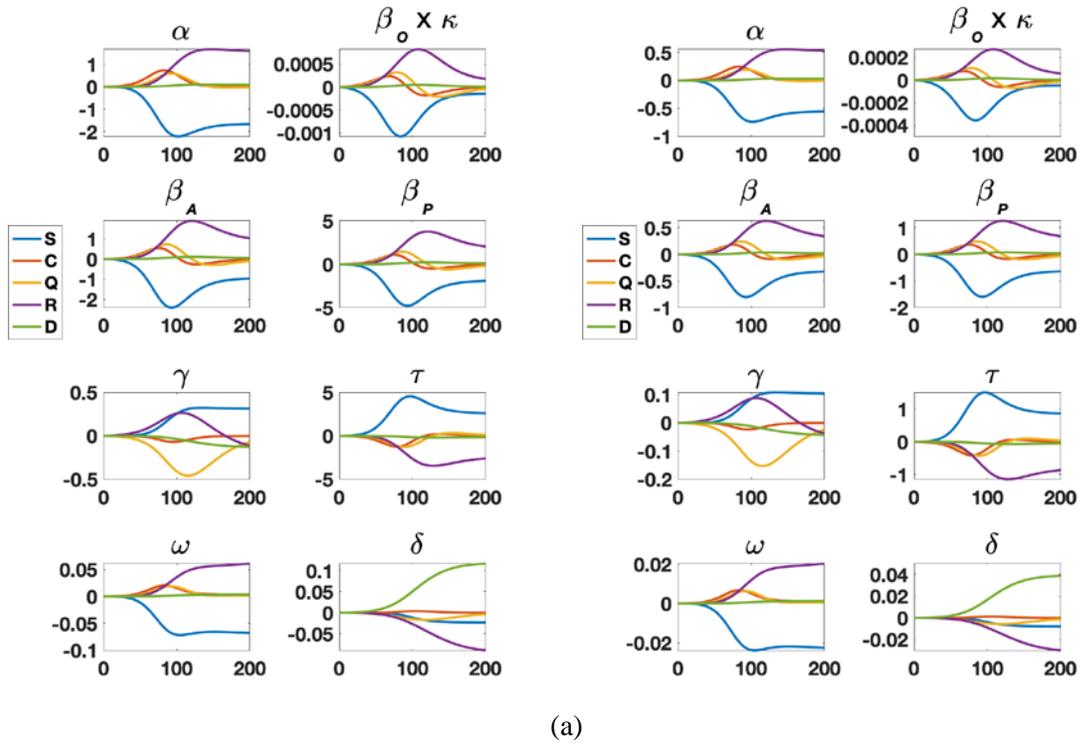

(a) (b)

Figure 2. (a) Local sensitivity analysis of the transmission model; (b) Local sensitivity analysis of the transmission model – Percentage on outcomes.

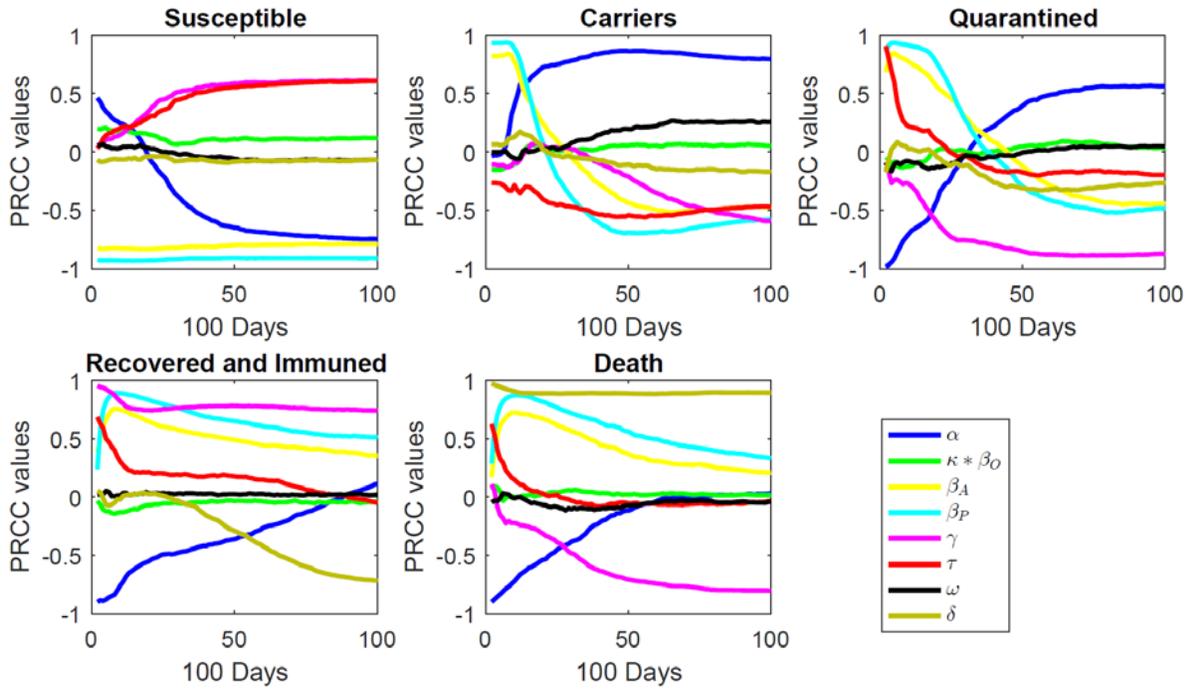

Figure 3. Global sensitivity analysis (PRCC) of the transmission model.

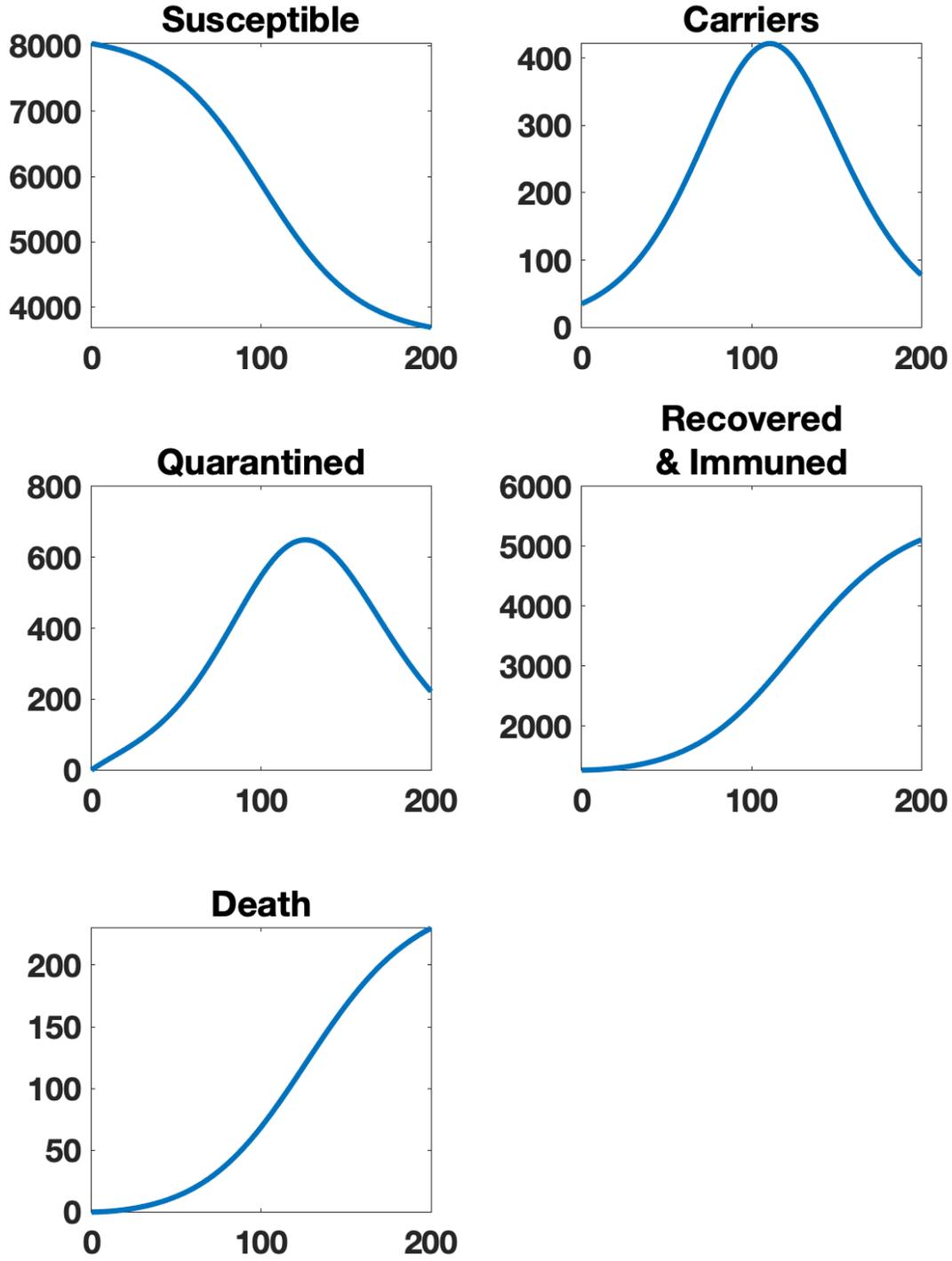

Figure 4. Outcome trajectories in the baseline scenario (Scenario I).

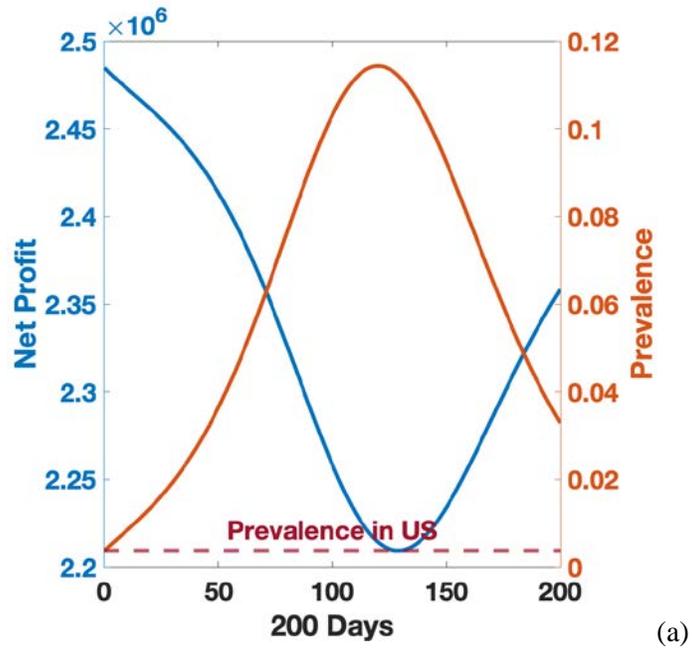

(a)

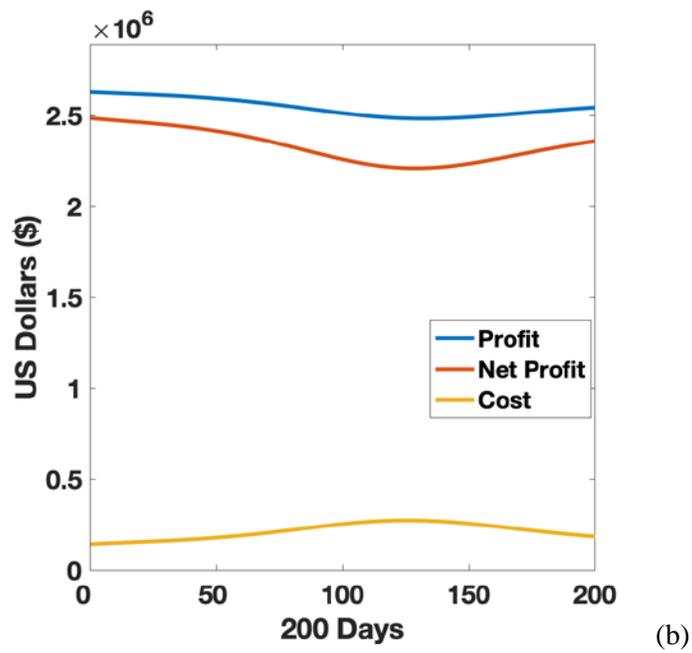

(b)

Figure 5. Scenario I. (a) Reopening feasibility based on both net profit and infection prevalence; (b) Cost, profit and net profit.

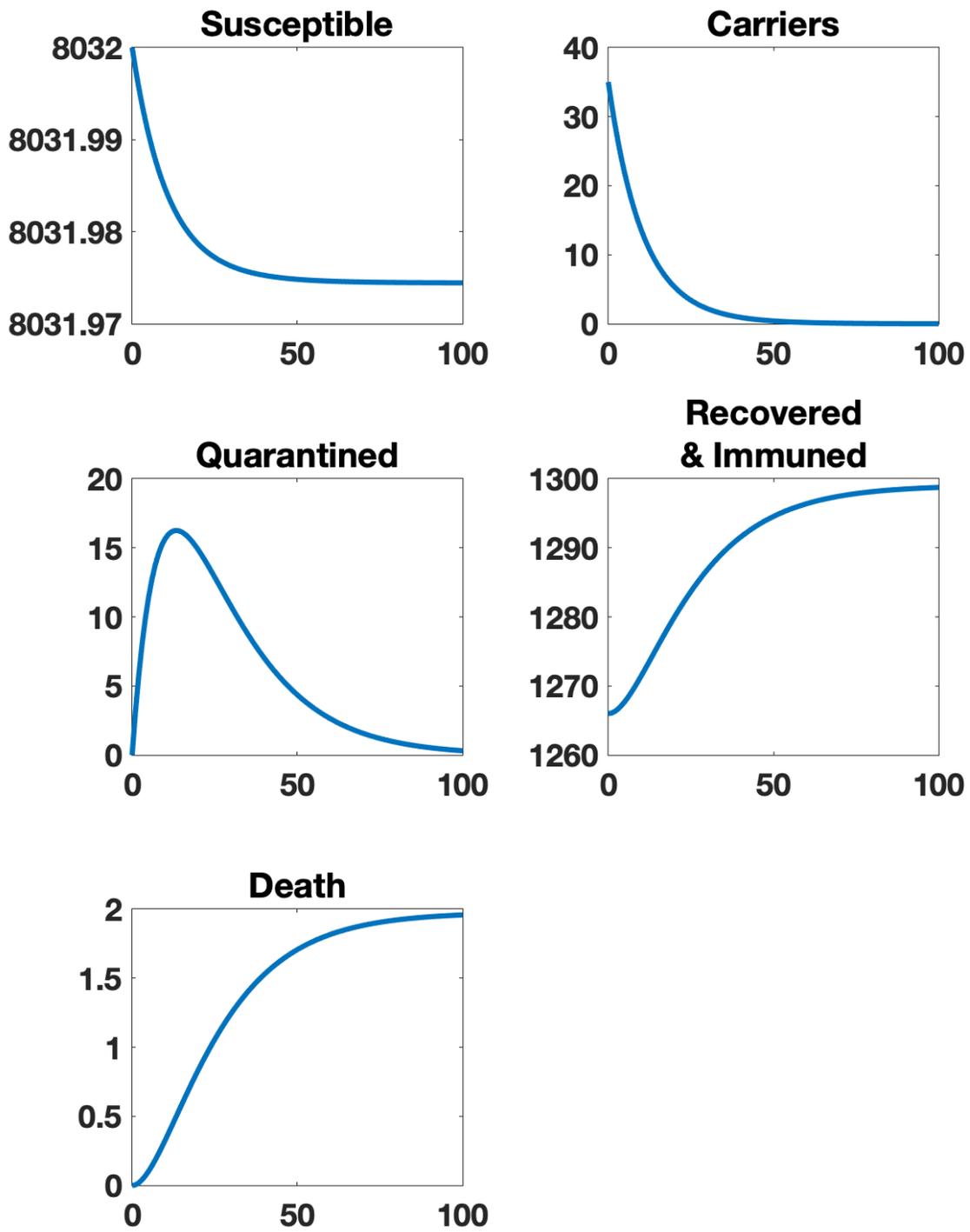

Figure 6. Outcome trajectories in Scenario II.

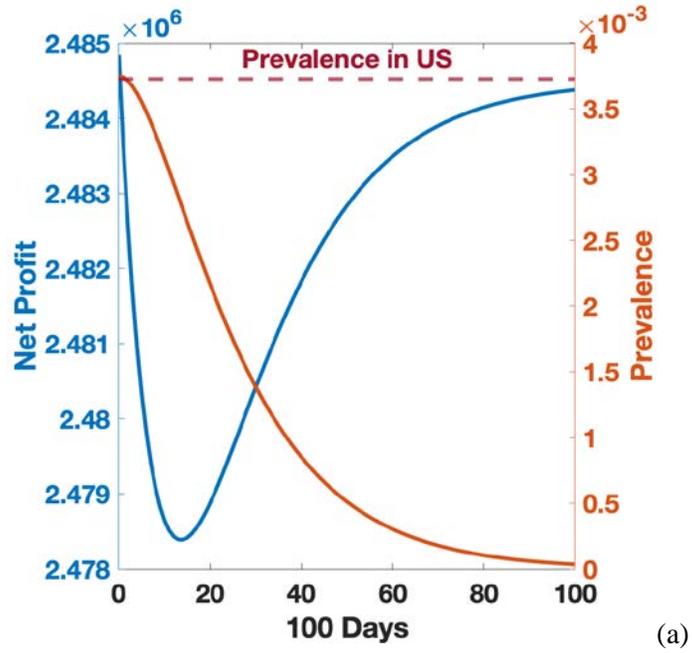
(a)

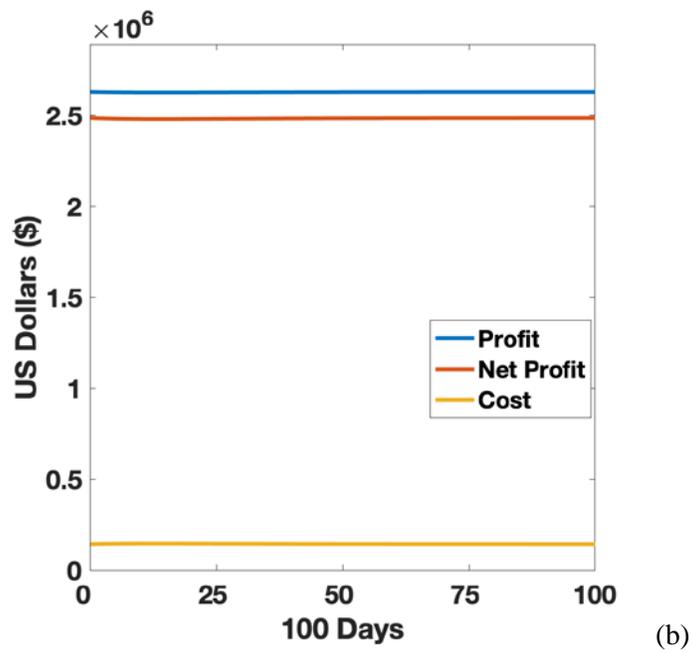
(b)

Figure 7. Scenario II. (a) Reopening feasibility based on both net profit and infection prevalence; (b) Cost, profit and net profit.

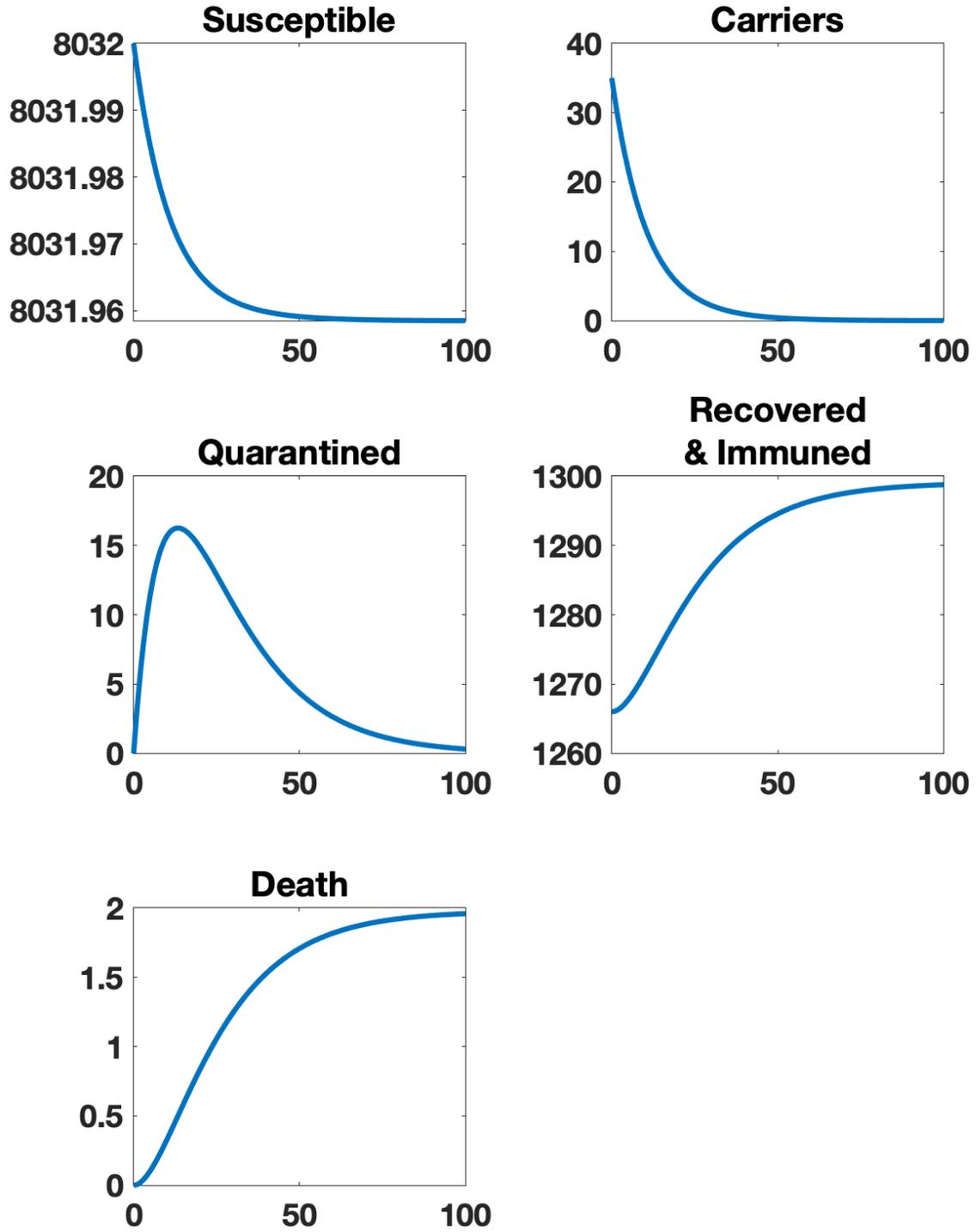

Figure 8. Outcome trajectories in Scenario III.

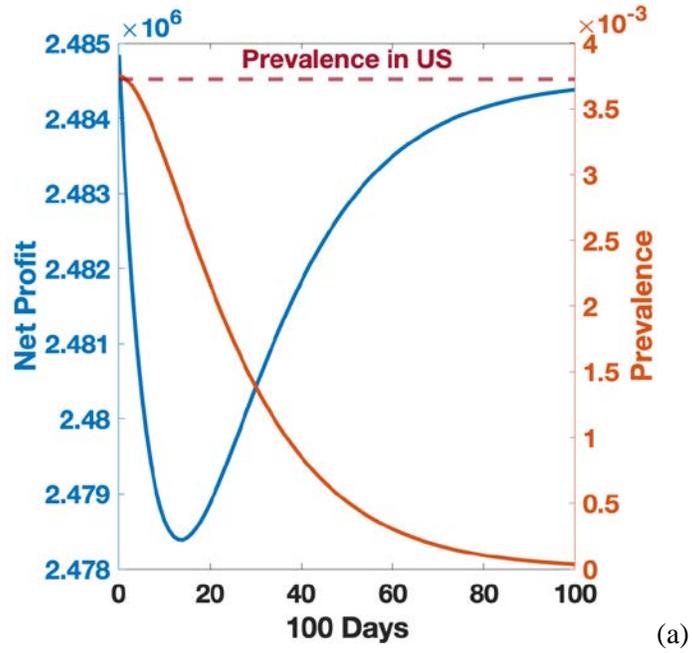
(a)

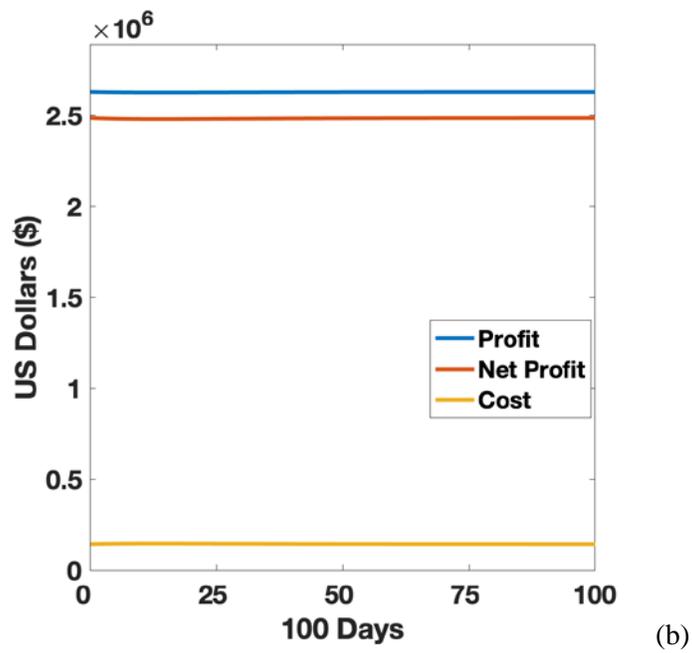
(b)

Figure 9. Scenario III. (a) Reopening feasibility based on both net profit and infection prevalence; (b) Cost, profit and net profit.

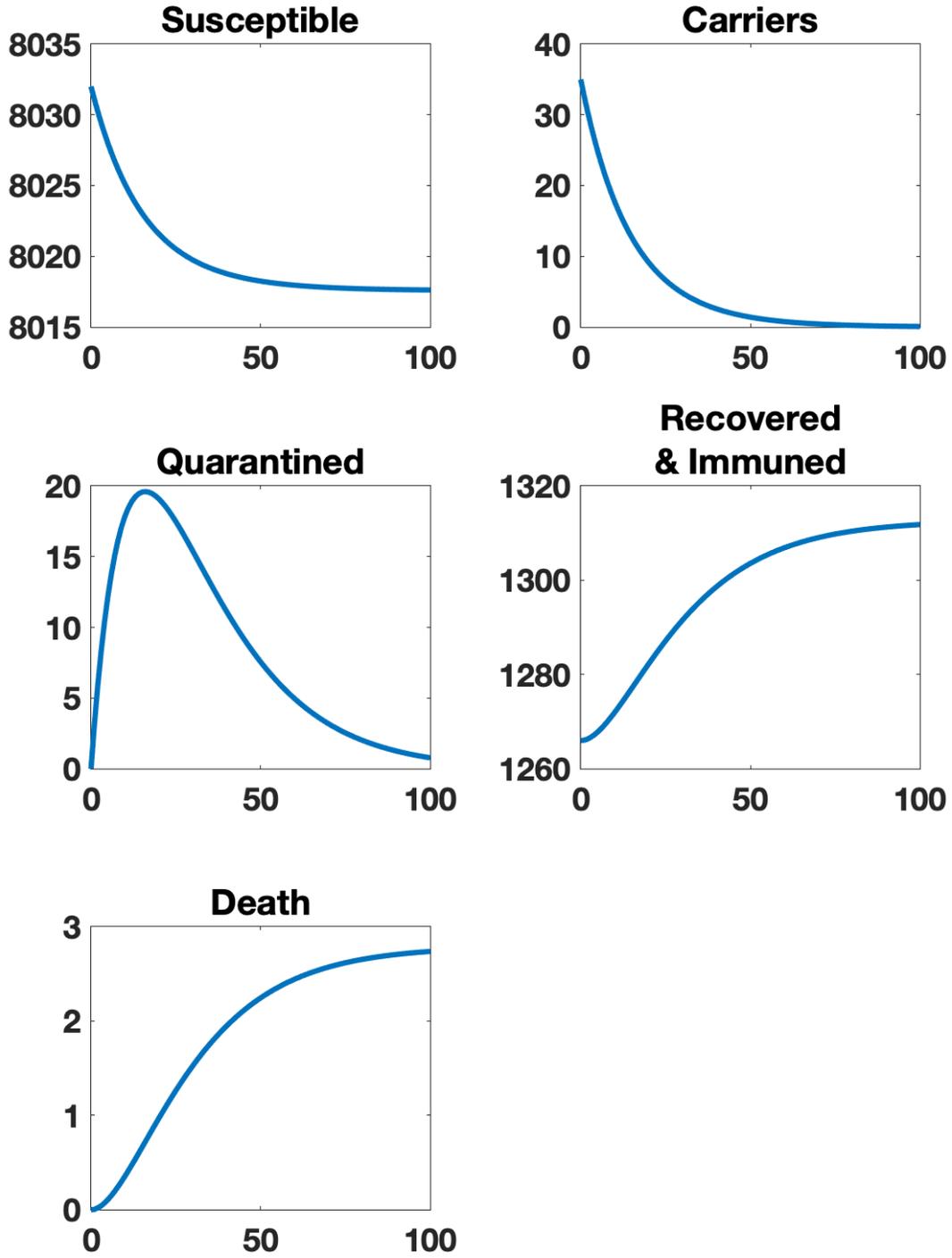

Figure 10. Outcome trajectories in Scenario IV.

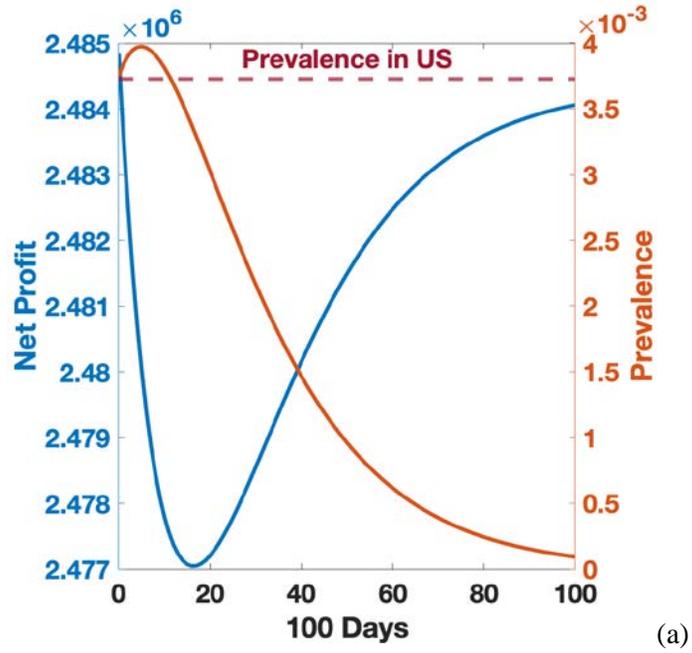

(a)

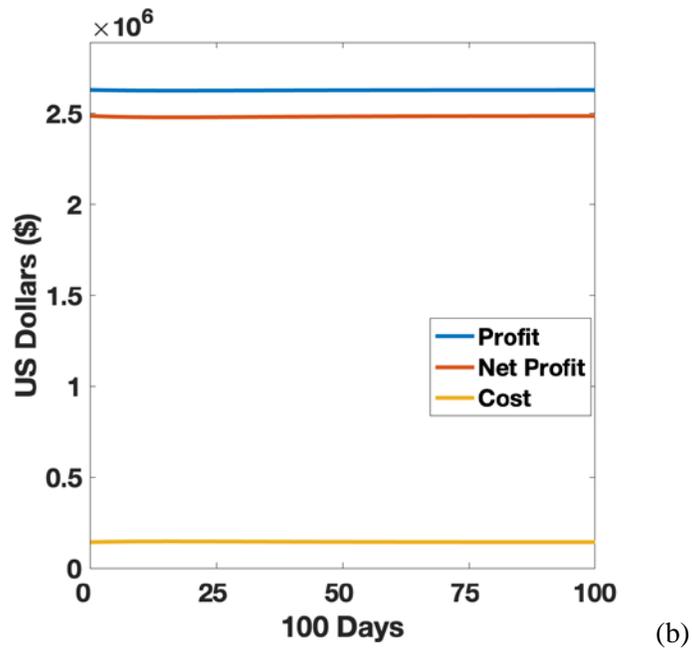

(b)

Figure 11. Scenario IV. (a) Reopening feasibility based on both net profit and infection prevalence; (b) Cost, profit and net profit.

Table 1. Variable and parameter definitions, values and sources.

| Notation | Definition | Unit | Value | Reference |
|---|---|---|---|---|
| $S$ | Susceptible | Persons | | |
| $C$ | Carrier | Persons | | |
| $Q$ | Quarantine | Persons | | |
| $R$ | Recovered | Persons | | |
| $D$ | Death | Persons | | |
| $N$ | Sum of S, C and R | Persons | | |
| $N_C$ | Number of employees in contact-based positions | Persons | | |
| $N_N$ | Number of employees in non-contact positions | Persons | | |
| $N_{NQ}$ | Number of employees in non-contact positions under quarantine | Persons | | |
| $\alpha$ | Proportion of asymptomatic carriers among all types of carriers | % | 50 (13.8, 75) | 15,26-30 |
| $\beta_O \cdot \kappa$ | The product of $\beta_O$ and $\kappa$ | $day^{-1}$ | $3.10 \times 10^{-7}$ ($2.42 \times 10^{-7}$, $3.88 \times 10^{-7}$) | 32 |
| $\kappa$ | Probability of one employee having contacts with infected carriers outside a business entity | % | NA | |
| $\beta_O$ | Infection rate associated with activities outside of the business entity | $day^{-1}$ | NA | |
| $\beta_A$ | Transmission rate of asymptomatic infection | $day^{-1}$ | 0.099 (0.0103, 0.814) | Calibrated from 14,44, 43 |
| $\beta_P$ | Transmission rate of presymptomatic and very-mildly-symptomatic infection | $day^{-1}$ | 0.197 (0.0333, 1.18) | Calibrated from 23, 25, 42 |
| $\tau$ | Rate of progression to symptomatic infection | $day^{-1} person^{-1}$ | 0.192 (0.143, 0.244) | Calibrated from 14,44 |
| $\omega$ | Rate of regression to carriage (e.g., even after treatment) | $day^{-1} person^{-1}$ | 0.00172 (0.000575, 0.00455 | Calibrated from 47,48 |
| $\gamma$ | Clearance rate of symptomatic infection, including the portion of negative test outcomes among exposed employees | $day^{-1}$ | 0.0535 (0.0178, 0.141) | 6,20 |
| $\delta$ | Death rate due to infection | $day^{-1}$ | 0.00320 | Calibrated from |

|  |  |  |  | (0.00013, 0.0167) | 25 |
|---|---|---|---|---|---|
| $\rho$ | Net profit per capita | $ per person per day | 400.73 | | 49 |
| $w$ | Average wage of employees | $ per person per da | 218.60 | | 51 |
|  |  |  |  |  |  |
| $\xi_1$ | Social distancing | % | 67.4 | | 12 |
| $\xi_2$ | Personal protective equipment (PPE) | $ per person per day | 3.60 | | 53-55 |
| $\xi_3$ | Routing sanitization | $ per environmental service staff per day | 10.45 | | 58 |
| $\xi_4$ | COVID test | $ per person per day | 0 | | assumed |
| $\xi_5$ | Non-contact thermometer | $ per person per day | (0.005, 0.01) | | 56,57 |
| $\xi_6$ | Case reporting and quarantine | $ per person per day | 0 | | |
| $\xi_7$ | Proportion of current hours of exposure to working environment | % | 70 (50, 100) | | |
| $\xi_8$ | Specific equipment for aerosol transmission (e.g., UVGI, HEPA filtration) | $ per person per day | 0.874 | | 54 |
| $\xi_9$ | WFH | $ per person per day | 0 | | assumed |
| $\xi_i$: Protection cost per person associated with the $i$-th Guideline; UVIG: ultraviolet germicidal irradiation; HEPA: high-efficiency particulate air filtration; ||||||

Table 2. Parameters value adjustment in different scenarios.

| | Scenario I (Baseline) | Scenario II | Scenario III | Scenario IV |
|---|---|---|---|---|
| Guidelines Implemented | None | All | 1,2,3,4,6,9 | 1,3,4,6,9 |
| $\beta_O \cdot \kappa$ (% × day$^{-1}$) | $3.100 \times 10^{-17}$ | $1.653 \times 10^{-20}$ | $2.684 \times 10^{-20}$ | $6.613 \times 10^{-18}$ |
| $\beta_A$ (day$^{-1}$) | 0.099 | $5.280 \times 10^{-05}$ | $8.571 \times 10^{-05}$ | $2.112 \times 10^{-02}$ |
| $\beta_P$ (day$^{-1}$) | 0.197 | $1.051 \times 10^{-04}$ | $1.706 \times 10^{-04}$ | $4.203 \times 10^{-02}$ |